\documentclass{aipproc}

\layoutstyle{8x11double}

\title{Astrometric Calibration and Estimate of the Systematic
Error in WXM Localizations Obtained by the Chicago Bayesian Method}

\author{C. Graziani}
{
address={
Department of Astronomy \& Astrophysics, University of Chicago,
5640 South Ellis Avenue, Chicago, IL 60637
},
email={carlo@oddjob.uchicago..edu},
}

\author{Y. Shirasaki}{
  address={JST/NASDA, Tsukuba, Ibaraki 305-8505, Japan}
}

\author{T. Donaghy}{
  address={Department of Astronomy \& Astrophysics, University of
Chicago, Chicago, IL 60637, USA}
}

\author{E. Fenimore}{
  address={Los Alamos National Laboratory, Los Alamos, NM 87545, USA}
}

\author{M. Galassi}{
  address={Los Alamos National Laboratory, Los Alamos, NM 87545, USA}
}

\author{N. Kawai}{
  address={Department of Physics, Tokyo Institute of Technology,
  Meguro-ku, Tokyo 152-8551, Japan},
  email={nkawai@hp.phys.titech.ac.jp}}

\author{D.Q. Lamb}{
  address={Department of Astronomy \& Astrophysics, University of
Chicago, Chicago, IL 60637, USA}
}

\author{T. Sakamoto}{
  address={Department of Physics, Tokyo Institute of Technology,
  Meguro-ku, Tokyo 152-8551, Japan}
}

\author{D. Takahashi}{
  address={Aoyama Gakuin University, Shibuya-ku, Tokyo 150-8366, Japan}
}

\author{T. Tamagawa}{
  address={RIKEN, Wako, Saitama 351-0198, Japan}
}

\author{T. Tavenner}{
  address={Los Alamos National Laboratory, Los Alamos, NM 87545, USA}
}

\author{K. Torii}{
  address={RIKEN, Wako, Saitama 351-0198, Japan}
}

\author{A. Yoshida}{
  address={Aoyama Gakuin University, Shibuya-ku, Tokyo 150-8366, Japan}
}

\author{R. Vanderspek}{
  address={MIT Center for Space Research, MIT, Cambridge, MA 02139, USA}
}

\begin{abstract}
WXM gives GRB localizations in instrument coordinates.  WXM
localizations must be converted to celestial coordinates using
spacecraft aspect information obtained by the optical cameras on HETE. 
We must therefore accurately determine the alignment of the WXM
boresight with respect to that of the optical cameras, in order to
accurately determine the celestial coordinates of WXM burst locations. 
We use a seven-parameter model that treats as free parameters the three
Euler angles of a pure rotation, two horizontal shifts of the
coded-aperture masks with respect to the detectors, and the heights of
the masks above the two detectors.  We determine the alignment by
fitting the model to a set of 252 WXM localizations of Sco X-1 obtained
between 23 April and 28 June 2001.  We estimate the systematic error in
WXM GRB locations by comparing the actual and the calculated locations
of Sco X-1.  We find that the systematic error corresponding to a
68.3\% confidence region is 1.7$'$, and the systematic error
corresponding to a 90\% confidence region is 2.4$'$. We find that this
astrometric solution also provides a satisfactory fit to an independent
sample of SGR and XRB events.  These results are consistent with the
astrometric calibration and the systematic error in WXM localizations 
derived independently using the RIKEN localization method.
\end{abstract}

\begin{document}

\maketitle

\section{Introduction}

The spacecraft placement of WXM is offset from that of the optical cameras
(the OPT subsystem).  The coordinate frames of WXM on of the optical
cameras are only aligned to within machining tolerances and thermal
shrinkage effects.  These misalignments are of the order of 0.4$^\circ$,
and must be corrected so that HETE can provide WXM locations accurate to
$5'-15'$.  Any systematic effects inherent in the localization procedure
compound the misalignment effects, and must also be corrected.

The HETE team employs two independent WXM localization pipelines, which
use different location algorithms.  The existence of two independent
WXM localization pipelines has been invaluable as a check of the
correctness of the design and implementation of each approach.  We
describe here the study of astrometric correction of the Chicago
(Bayesian) location algorithm \citep{graziani02}.  The calibration of
the RIKEN location algorithm is described elsewhere in these
proceedings.\citep{shirasaki02}

We have performed this astrometric calibration on-orbit, using  sources
whose locations are accurately known --- Sco~X-1, several X-Ray Burst
(XRB) sources, and two Soft-Gamma Repeaters (SGRs).

\begin{figure}[t]
\resizebox{0.85\textwidth}{!}{
\includegraphics[scale=0.5]{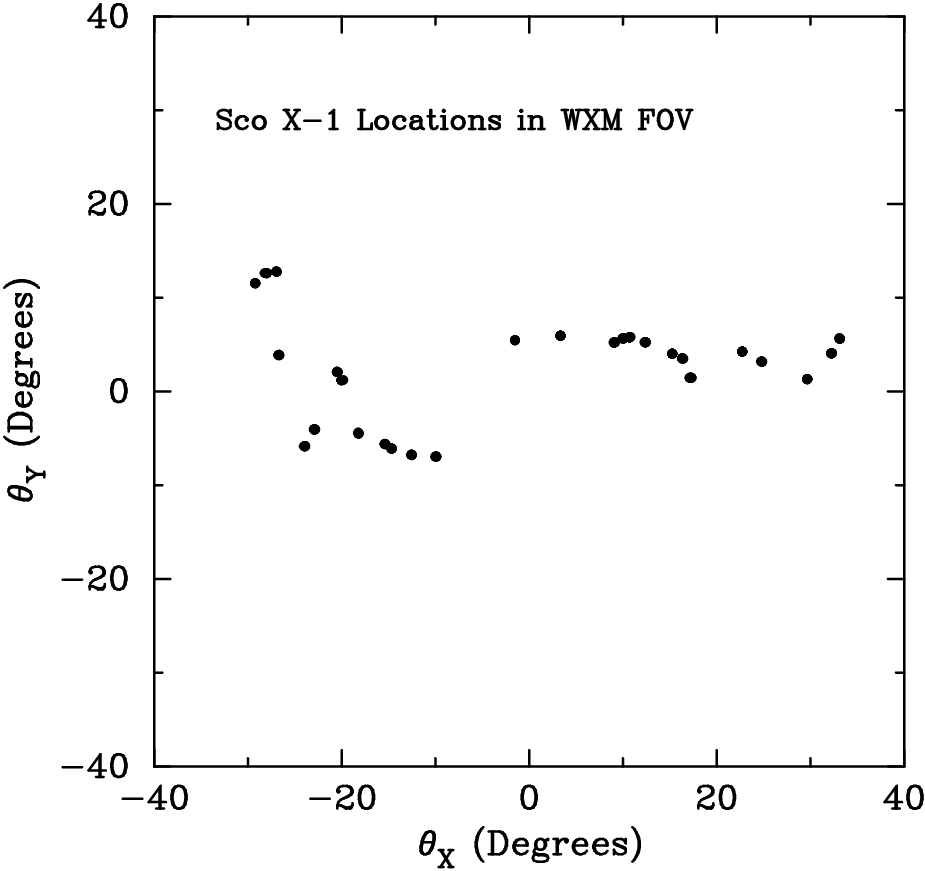}
\hspace{0.2in}
\includegraphics[scale=0.5]{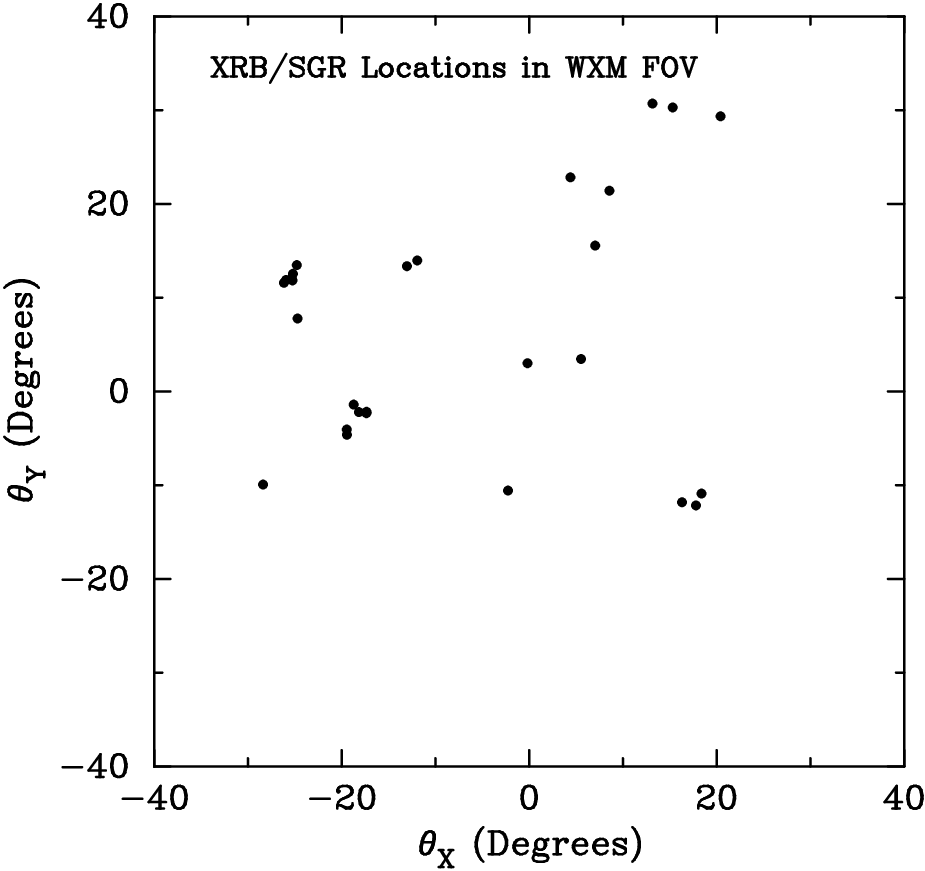}
}

\caption{Location of the calibration sources in the WXM FOV.  Left
Panel:  Location of Sco~X-1 during each of the 27 health check sessions
used for astrometric calibration.  Right panel:  Locations of 28
XRB and SGR sources.} 
\label{source_fov}

\end{figure}

\section{The Data}

The calibration was performed using ``RAW'' data of Sco~X-1 obtained in
the course of the daily health check observations made during the
period 2001~April~23 and 2001~June~28.  There are 252 Sco~X-1 locations
determined using 27 sessions of 180s each.  The statistical errors for
these locations are estimated from the scatter in the locations
independently for each session.  The location of Sco~X-1 in the WXM
FOV for each of these 27 sessions is shown in the left panel of Figure
(\ref{source_fov}).

We also derived the locations of 23 XRBs from known XRB sources, of
three SGR1900+14 bursts, and of two SGR1806-20 bursts.  We used these
locations as an independent check on the quality of the fit to the
Sco~X-1 data.  The statistical errors for these locations are produced
directly by the Bayesian location process.  Their locations in the WXM
FOV are shown in the right panel of Figure (\ref{source_fov}).

\section{The Astrometric Model}

The transformation from WXM to OPT coordinates is assumed to have
the following form:

\begin{equation}
\tan\left(\theta_X^{(True)}\right)=
S_X \tan\left(\theta_X^{(Meas)}\right) + d_X,
\end{equation}
\begin{equation}
\tan\left(\theta_Y^{(True)}\right)=
S_Y \tan\left(\theta_Y^{(Meas)}\right) + d_Y,
\end{equation}
\begin{equation}
\vec n^{(Opt)}=
{\bf M}(\epsilon_1,\epsilon_2,\epsilon_3)
\cdot\vec n^{(WXM)}\left(\theta_X^{(True)},\theta_Y^{(True)}\right).
\end{equation}

The projection angles $\theta_X^{(Meas)}$ and $\theta_Y^{(Meas)}$ are the
angles that result from the Bayesian location analysis.  The model
corrects these angles to produce ``true'' projection angles
$\theta_X^{(True)}$ and $\theta_Y^{(True)}$.  The unit vector $\vec
n^{(WXM)}$ is the direction vector to the source defined by the projection
angles $\theta_X^{(True)}$, $\theta_Y^{(True)}$.  The unit vector $\vec
n^{(Opt)}$ is the direction vector to the source in the OPT frame.

${\bf M}$ is a rotation matrix.  $\epsilon_1$, $\epsilon_2$, and
$\epsilon_3$ are Euler angles.  $S_X$ and $S_Y$ are scale parameters that
can represent unknown changes in the height of the coded aperture masks.
$d_X$ and $d_Y$ are shifts that can represent unknown mis-alignments of
the masks with respect to the detectors.  The model thus has 7 free
parameters.

\section{Sco X-1 Calibration Results}

With no astrometric correction, the $\chi^2$ of the fit is $2.9\times
10^5$ for 504 DOF, and the RMS deviation (computed versus actual
locations) is 0.39$^\circ$.

\begin{figure}[h]
\begin{minipage}[t]{3.0in}
\includegraphics[scale=0.45]{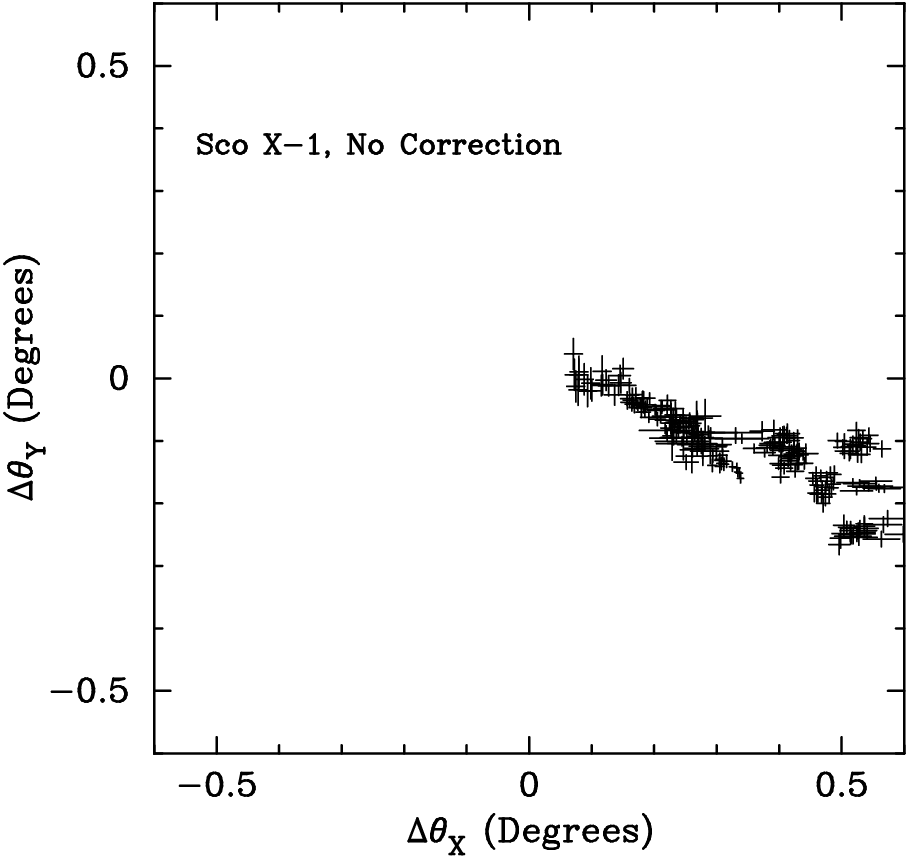}
\\
\includegraphics[scale=0.45]{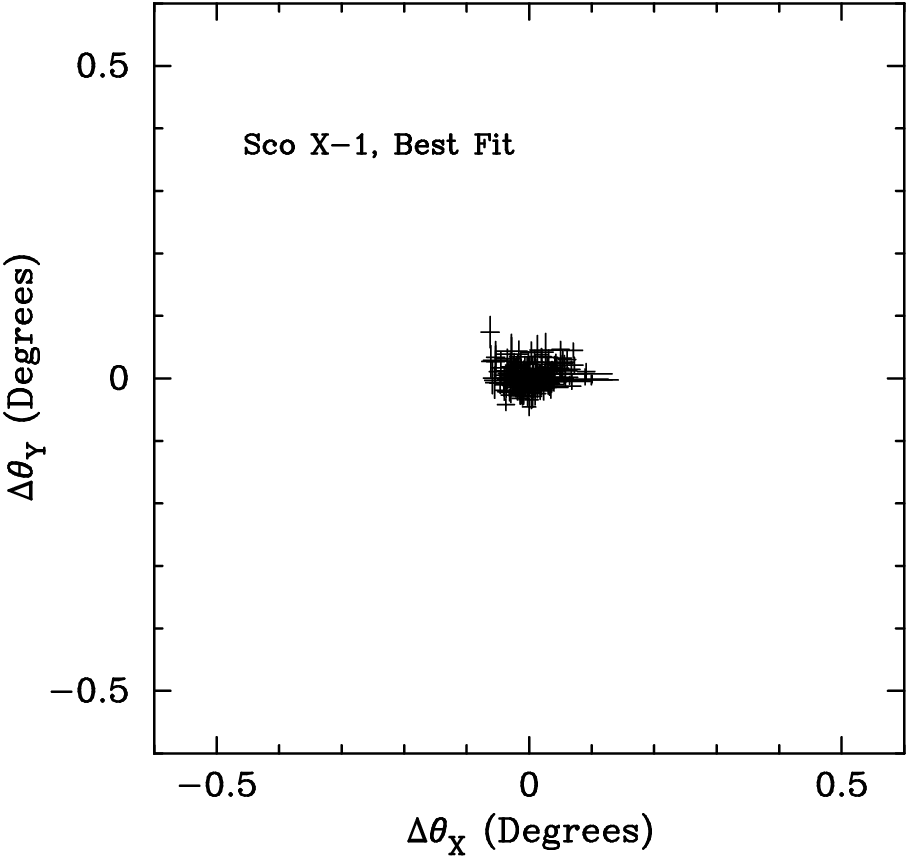}
\end{minipage}

\caption{Deviations of derived locations of Sco X-1 from true source location.  Top
Top:  Assuming no astrometric correction.  Bottom:  Assuming the
astrometric correction that best fits the Sco X-1 data. 
}
\label{sco_dev}
\end{figure}

The deviations from the true locations are shown in the left panel of
Figure (\ref{sco_dev}), together with the estimated location errors,
which are typically in the 1$'$-2$'$ range.  The misalignment of the WXM
and OPT frames is clearly manifested in the Figure.

The best-fit astrometric correction results in a $\chi^2=1517$ for 497
DOF.  The fit gives an RMS deviation of 2.0'.  The deviations from the
true locations are shown in the right panel of Figure (\ref{sco_dev}).

While clearly a distinct improvement over the null correction fit, the
quality of this fit is rather poor --- the $Q$-value for $\chi^2=1517$
from the $\chi^2$ distribution with 497 DOF is $\sim 10^{-103}$.  This
excess $\chi^2$ is attributable to systematic error.  The source of the
error is presumably a compounding of the inadequacy of the astrometric
model with the limitations of the location analysis.

We estimate the magnitude of the systematic error as follows: we add
(in quadrature) a systematic error to the statistical error so as to
bring the $\chi^2$/DOF down to about 1.  In this way we find that the
systematic error corresponding to a 68.3\% confidence region is 1.7$'$,
and the systematic error corresponding to a 90\% confidence region is
2.4$'$.  These results are consistent with those found using the RIKEN
location algorithm.\citep{shirasaki02}

\section{An Independent Check:  The SGR/XRB Sample}

When the astrometric model that best fits the Sco~X-1 data is applied
to the XRB/SGR bursts and the derived locations are compared with the
known locations of the sources, the result is $\chi^2=39.6$ for 56 DOF
(P=0.05).  The RMS deviation is 7.0'.  The Sco~X-1 astrometric solution
thus provides a satisfactory fit to this independent sample of event
locations.  No systematic error is needed in this fit, because these
sources are less bright --- and thus less accurately located --- than
Sco~X-1.  The typical statistical error for this sample is about 6$'$.

\begin{figure}[t]
\begin{minipage}[t]{3.0in}
\includegraphics[scale=0.45]{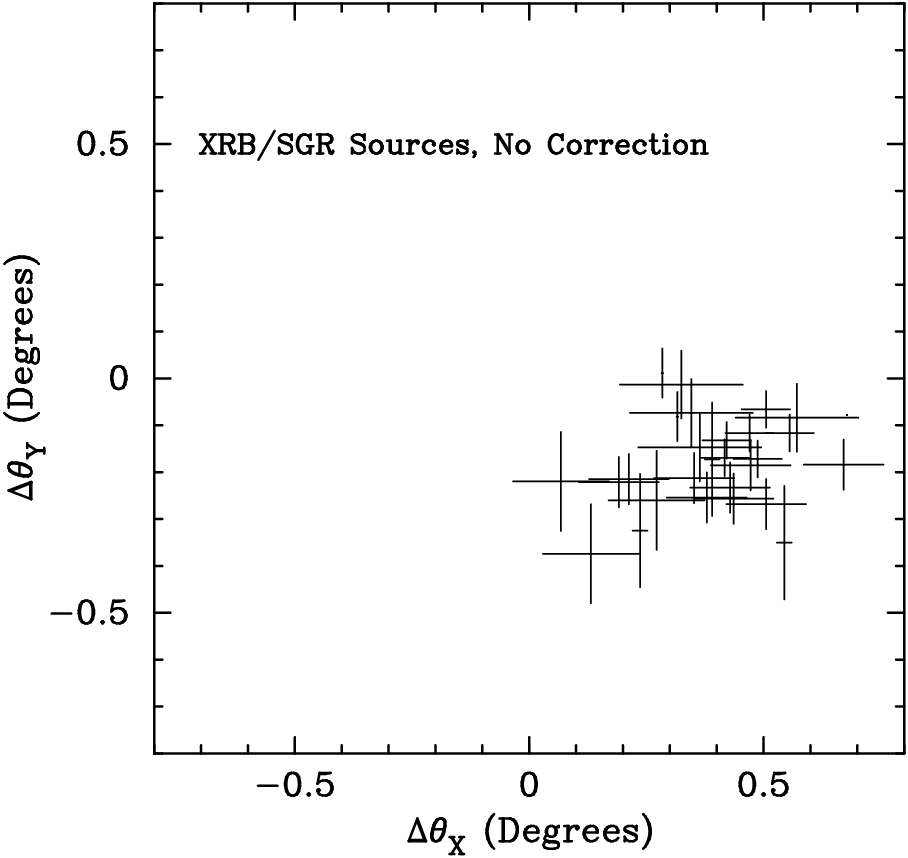}
\\
\includegraphics[scale=0.45]{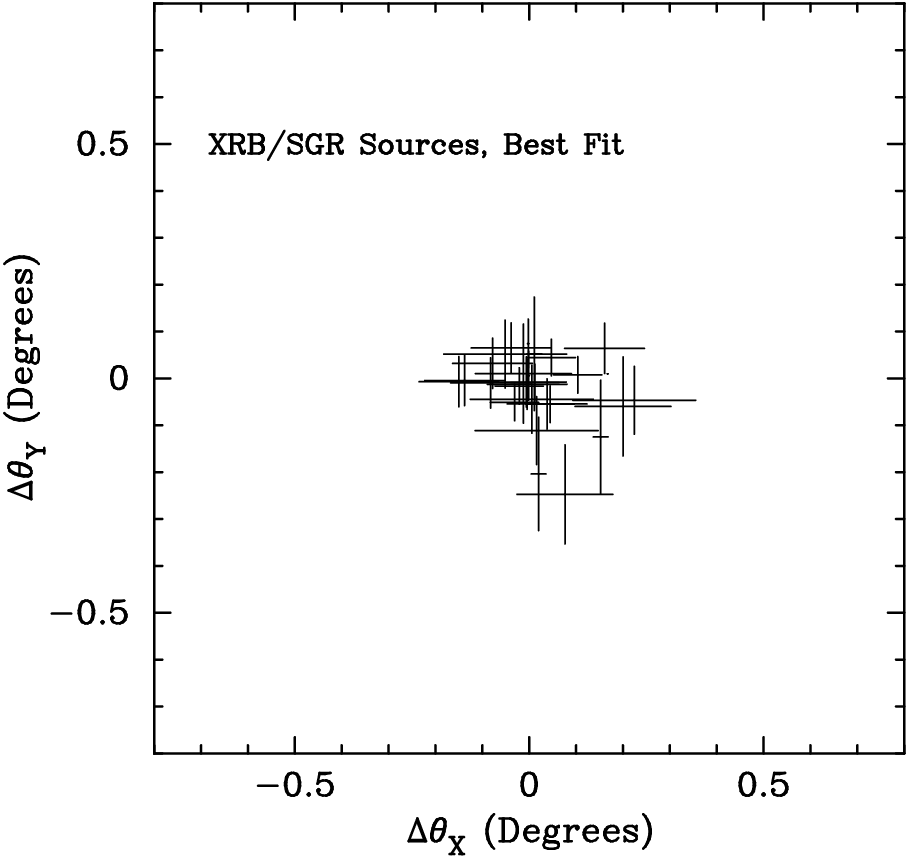}
\end{minipage}

\caption{Deviations of derived locations of XRB/SGR sample from true
source location.  Top:  Assuming no astrometric correction.  Bottom: 
Assuming the astrometric correction that best fits the Sco X-1 data. 
}
\label{xrb_sgr_dev}
\end{figure}

\end{document}